\documentclass[twocolumn]{aastex62}

\usepackage{color}
\usepackage{graphicx}
\usepackage[title]{appendix}
\usepackage{diagbox}
\usepackage{nicefrac}
\usepackage{amsmath}
\usepackage{csvsimple}
\usepackage{longtable}
\usepackage{rotating}
\usepackage{bm}

\newcommand{\teff}{$T_{\rm eff}$}
\newcommand{\logg}{$\log g$}
\newcommand{\feh}{[Fe/H]}
\newcommand{\s}{\emph{S}}

\newcommand{\figu}[1]{Figure~\ref{#1}}
\usepackage{nicefrac}
\usepackage{multirow}
\shorttitle{reconstructing stellar noise}
\shortauthors{Jinghua Zhang et al.}

\begin{document}

\title{Reconstructing  Intrinsic Stellar Noise with Stellar Atmospheric Parameters and Chromospheric Activity}

\author[0000-0002-2510-6931]{Jinghua Zhang}
\affiliation{South-Western Institute for Astronomy Research, Yunnan University, Chenggong District, Kunming 650500, China}
\affiliation{CAS Key Laboratory of Optical Astronomy, National Astronomical Observatories, Chinese Academy of Sciences, Beijing 100101, China}

\author{Maosheng Xiang}
\affiliation{CAS Key Laboratory of Optical Astronomy, National Astronomical Observatories, Chinese Academy of Sciences, Beijing 100101, China}
\affiliation{Institute for Frontiers in Astronomy and Astrophysics, Beijing Normal University,  Beijing 102206, China}
\email{msxiang@nao.cas.cn}

\author{Jie Yu}
\affiliation{Max Planck Institute for Solar System Research, G\"{o}ttingen, D-37077, Germany} 
\email{jie.yu@anu.edu.au}

\author{Jian Ge}
\affiliation{Shanghai Astronomical Observatory, Chinese Academy of Sciences, Shanghai 200030, China}

\author{Ji-Wei Xie}
\affiliation{School of Astronomy and Space Science, Nanjing University, Nanjing 210023, People’s Republic of China}
\affiliation{Key Laboratory of Modern Astronomy and Astrophysics, Ministry of Education, Nanjing 210023, China}

\author[0000-0003-3491-6394]{Hui Zhang}
\affiliation{Shanghai Astronomical Observatory, Chinese Academy of Sciences, Shanghai 200030, China}

\author{Yaguang Li}
\affiliation{Sydney Institute for Astronomy (SIfA), School of Physics, University of Sydney, NSW 2006, Australia}

\author{You Wu}
\affiliation{National Astronomical Observatories, Chinese Academy of Sciences, Beijing 100101, China}

\author{Chun-Qian Li}
\affiliation{CAS Key Laboratory of Optical Astronomy, National Astronomical Observatories, Chinese Academy of Sciences, Beijing 100101, China}
\affiliation{School of Astronomy and Space Science, University of Chinese Academy of Sciences,  Beijing 100101, China}

\author{Shaolan Bi}
\affiliation{Department of Astronomy, Beijing Normal University, Beijing 100875, China}
\affiliation{Institute for Frontiers in Astronomy and Astrophysics, Beijing Normal University,  Beijing 102206, China}

\author[0000-0002-8609-3599]{Hong-Liang Yan}
\affiliation{CAS Key Laboratory of Optical Astronomy, National Astronomical Observatories, Chinese Academy of Sciences, Beijing 100101, China}
\affiliation{School of Astronomy and Space Science, University of Chinese Academy of Sciences,  Beijing 100101, China}
\affiliation{Institute for Frontiers in Astronomy and Astrophysics, Beijing Normal University,  Beijing 102206, China}

\author[0000-0002-0349-7839]{Jian-Rong Shi}
\affiliation{CAS Key Laboratory of Optical Astronomy, National Astronomical Observatories, Chinese Academy of Sciences, Beijing 100101, China}
\affiliation{School of Astronomy and Space Science, University of Chinese Academy of Sciences,  Beijing 100101, China}
\email{sjr@nao.cas.cn}




\begin{abstract}
Accurately characterizing intrinsic stellar photometric noise induced by stellar astrophysics, such as stellar activity, granulation, and oscillations, is of crucial importance for detecting transiting exoplanets. In this study, we investigate the relation between the intrinsic stellar photometric noise, as quantified by the $Kepler$ rrmsCDPP measurement, and the level of stellar chromospheric activity, as indicated by the \s{}-index of Ca~{\sc ii} H\,K lines derived from the LAMOST spectra. Our results reveal a clear positive correlation between \s{}-index and rrmsCDPP, and the correlation becomes more significant at higher activity levels and on longer timescales. We have therefore built an empirical relation between rrmsCDPP and \s{}-index as well as \teff{}, \logg{}, \feh{}, and apparent magnitude with the \texttt{XGBoost} regression algorithm, using the LAMOST-\emph{Kepler} common star sample as the training set. This method achieves a precision of $\sim$20~ppm for inferring the intrinsic noise from the \s{}-index and other stellar labels on a 6-hour integration duration. We have applied this empirical relation to the full LAMOST DR7 spectra database, and obtained the intrinsic noise predictions for 1,358,275 stars. The resultant catalog is publicly available and expected to be valuable for optimizing target selection for future exoplanet-hunting space missions, such as the Earth 2.0 mission.
\end{abstract}


\section{Introduction} \label{sec:intro}
Activities caused by surface magnetism are prevalent on stellar surfaces, which provides us probe to study the dynamo mechanism of stellar interiors and further contribute to our understanding of their roles playing in stellar evolution \citep[see][for a review]{Charbonneau2023}. However, magnetic activities hold negative influences in the exploration of exoplanets, especially in the search for Earth-analog exoplanets, and further in exoplanetary system environments \citep[e.g.][]{Cegla2019, Hatzes2019}. Activity-induced radial velocity variations can mimic or conceal the Doppler signatures of orbiting planets, resulting in difficult even false detections of exoplanets with the Doppler method \citep[see ][for a review]{Meunier2021}. With the transiting method to search for exoplanets, stellar activities are also found to be one of the sources in the higher than expected noise of the \emph{Kepler} photometric time series \citep{Gilliland2011ApJS, Gilliland2015AJ}. 

Given that the non-stationary noise has impact on the detectability of the transit signature of the candidate, the noise levels over the 14 integration duration (i.e. [1.5, 2.0, 2.5, 3.0, 3.5, 4.5, 5.0, 6.0, 7.5, 9.0, 10.5, 12.0, 12.5, 15.0] hr) are determined by \emph{Kepler}’s transiting planet search (TPS) pipeline module \citep{Jenkins2010SPIE, Tenenbaum2012ApJS} for each light curve in the \emph{Kepler} photometric time series. The noise metrics are referred to as the combined differential photometric precision (CDPP), which are intended to be either the observed noise in a specified temporal domain or the predicted noise level in the same temporal domain from rolling up all contributing factors\citep[see][for detailed definition]{Christiansen2012PASP}. The early \emph{Kepler} on-orbit results \citep[e.g.][]{Christiansen2010} showed that the CDPP at the nominal 6.5 hr (one-half the duration for a central transit of a true Earth-analog) for solar-type stars of 12th in the \emph{Kepler} band magnitude are 30 parts per million (ppm), which are commonly 50\% higher than expected in the initial plan \citep{Jenkins2002ApJ}. The transit signals of earth-sized planets and smaller ones are thereby probably heavily drowned in noise.

It is necessary to perform an in-depth analysis of the noise properties, from which we may achieve a better understanding of the stellar noise and facilitate future planet-search missions. The properties of \emph{Kepler} photometric noise have been studied systematically \citep[e.g.][]{Gilliland2011ApJS, Gilliland2015AJ}. By analyzing the early release of data from Quarters 2 to 6 in 2009–2010, \citet{Gilliland2011ApJS} showed that the \emph{Kepler} observed noise can be decomposed into a few terms: fundamental terms (Poisson and readout noise), added noise due to the instrument and that intrinsic to the stars. Among them, the intrinsic stellar noise mainly due to stellar activity turned out to be the major contributor to CDPP, which strongly deviates from expectations since this term is twice the budgeted value \citep{Jenkins2002ApJ}. Considering data spanning 4 years, \citet{Gilliland2015AJ} revisited a similar analysis of the noise observed by \emph{Kepler}. On one hand, they found that the instrumental noise levels have dropped with the inclusion of more data, particularly the ones that have been processed recently. On the other hand, they showed that the intrinsic stellar noise levels have remained almost unchanged with the adoption of the newer data release, which is reasonable since updates to the pipeline cannot remove the intrinsic stellar noise.

There are still unanswered questions over how would stellar activity be related to photometric noise, or rather, to intrinsic stellar noise? And, is there a way to predict photometric noise and intrinsic stellar noise level based on the level of stellar activity? The answers to these questions are crucial for prioritizing targets in future Earth-analog exoplanet searches, such as the Earth 2.0 (ET) space mission \citep{Ge2022arXiv,ge2022spie,zhang2022spie}, a proposed Chinese space mission to detect thousands of small/low-mass exoplanets over a wide range of orbital periods. With the spectroscopic observations by the LAMOST survey \citep{Zhao2006, Cui2012, DeCat2015, Yan2022}, we can measure stellar chromospheric activity for millions of stars. This will facilitate us to investigate the dependence of photometric noise on stellar activity and further tackle these questions.

In this work, we focus on FGK-type stars to investigate the impact of chromospheric activity on intrinsic noise. We then predict the noise by accounting for chromospheric activity as well as stellar fundamental parameters for stars observed by the entire LAMOST survey. In Section \ref{sec:data}, we explain the sample selection. In Section \ref{sec:distribution}, we analyze the relation between stellar activity levels and stellar intrinsic noise for solar-type stars. We predict noise level for stars in the LAMOST field using machine learning algorithms in Section \ref{sec:fitting noise by ML}. Section \ref{sec:Discussion} and Section \ref{sec:conclusions} are discussion and conclusion, respectively.

\section{stellar samples} \label{sec:data}
We use the Kepler Stellar Properties Catalog for Q1-Q17 DR25 Transit Search which includes robust rms of the CDPP (hereafter, rrms CDPP) values over different integration durations from 1.5 to 15 hours \citep{Mathur2016ksci}. The one-half the duration for a central transit of a true Earth-analog is nearly 6.5 hours. Accordingly, we divide the accessible 6-hr rrms CDPPs by $(13/12)^{0.5}$ = 1.041 to approximate 6.5-hr rrms CDPPs.

\begin{figure}
\includegraphics[scale=0.45]{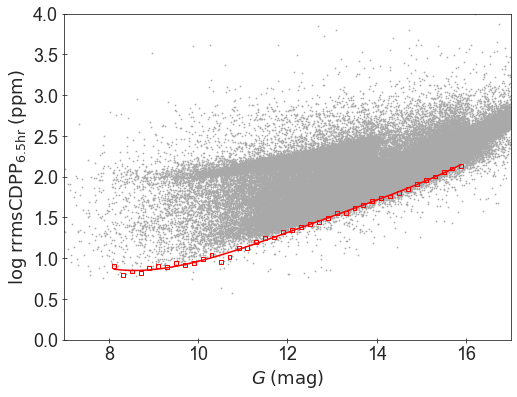}
\centering
\caption{Observed rrms CDPP on 6.5-hr timescale as a function of stellar apparent magnitude in the Gaia \emph{G} bandpass for \emph{Kepler} stars. The red square symbols at the lower envelope represent the bottom 0.5-th percentile of points in bins of 0.2 magnitudes wide. The red curve represents a 4th-order polynomial fit to the points shown by red squares, which serves as a lower border of the rrms CDPP as a function of magnitude.
\label{fig:cdpp06p5_vs_Gmag}}
\end{figure}
\begin{deluxetable*}{cccccccccccc}
\tiny
\tablecaption{The coefficients of the polynomial terms for the Poisson noise on different timescales. \label{tab:fit}}
\tablehead{
  & $\mathrm{CDPP}_{1.5\mathrm{h}}$ & $\mathrm{CDPP}_{2.5\mathrm{h}}$ & $\mathrm{CDPP}_{3.5\mathrm{h}}$ & $\mathrm{CDPP}_{5.0\mathrm{h}}$ & $\mathrm{CDPP}_{6.0\mathrm{h}}$ & $\mathrm{CDPP}_{6.5\mathrm{h}}$ & $\mathrm{CDPP}_{7.5\mathrm{h}}$ &  $\mathrm{CDPP}_{9.0\mathrm{h}}$ & $\mathrm{CDPP}_{10.5\mathrm{h}}$ & $\mathrm{CDPP}_{12.0\mathrm{h}}$ & $\mathrm{CDPP}_{15.0\mathrm{h}}$ }
\startdata
$p_0$ & 0.000414& 0.000534& 0.000507& 0.000628& 0.000646& 0.000646& 0.000627& 0.000631& 0.000591& 0.000550& 0.000447 \\
$p_1$& -0.020747& -0.026850& -0.025847& -0.031968& -0.033076& -0.033076& -0.032387& -0.032608& -0.030626& -0.028575& -0.023437 \\
$p_2$& 0.391895& 0.508094& 0.496268& 0.611070& 0.635465& 0.635465& 0.627501& 0.632242& 0.596405& 0.559023& 0.464380 \\
$p_3$& -3.106706& -4.083044& -4.050601& -4.998701& -5.231168& -5.231163& -5.210254& -5.258316& -4.977679& -4.682999& -3.927053 \\
$p_4$& 9.652102& 12.633736& 12.711457& 15.574175& 16.361260& 16.343797& 16.418002& 16.583189& 15.767899& 14.908123& 12.679827 \\
\enddata
\end{deluxetable*}

We derive stellar parameters, effective temperature (\teff{}), surface gravity (\logg{}) and metallicity (\feh{}) from the LAMOST DR7 low-resolution spectra \footnote{\url{http://dr7.lamost.org}} using the data-driven Payne ({\sc DD-Payne}) \citep{Xiang2019ApJS}. For our purposes, we restrict attention to FGK-type stars with \teff{} in the range $3800\sim6500$\,K. Stars hotter than 6500 K that generally fall into the classical instability strip are excluded. In addition, both eclipsing binaries \citep{Kirk2016AJ} and \emph{Kepler} objects of interest hosting planet candidates being flagged in DR25 \citep{Thompson2018ApJS} are excluded. 

To derive the activity proxy, i.e. \s{}-index of Ca {\sc ii} HK lines, we then consider stars with signal-to-noise ratios (S/Ns) of the LAMOST spectra higher than 50 to place a lower limit on the quality of the spectroscopic observations. Our final sample includes 39,056 stars. Following methods described in \citet{Zhang2020ApJS}, we calculate \s{}-index as the ratio of the integrated fluxes in the cores of Ca {\sc ii} H and K lines to that in the nearby pseudo-continuum. To perform the integration, we use a triangle function with a full width at half-maximum (FWHM) of 1.09\,\AA{} centered at 3968\,\AA{} and 3934\,\AA{} for H and K lines, respectively. While for the nearby pseudo-continuum, we use a rectangular function with a width of 20\,\AA{} centered at 4001\,\AA{} and 3901\,\AA{}. 

Repeat measurements for common stars in LAMOST observations can provide a good estimate for the internal precision of our \s{}-index measurements. In our sample, there are 5837 stars that have been observed by LAMOST at least three visits. For 84\% of these stars, the $S$-index scatter among individual measurements is below 0.02 dex. To better understand possible systematics in the $S$-index measurements, we have also implemented an external comparison with literature $S$-index from high-resolution spectra. A detailed discussion about the results of the external comparison is presented in Sect.~\ref{sec:Discussion}. Briefly, both a systematic trend and considerable scatter are present between our measurements and literature. However, as the effectiveness of our data-driven method of inferring stellar intrinsic noise is mostly determined by internal consistency rather than the absolute scale of the $S$-index measurements, we expect our results presented in the current work to be robust, given the small internal errors in our \s{}-index measurements.
\section{noise dependence on stellar activity} \label{sec:distribution}
\subsection{determining intrinsic stellar noise} \label{sec:intrinsic noise}
In order to quantify the intrinsic stellar noise of stars, we first subtract the photon noise and readout noise from the CDPP. It has been shown that the lower bound on the distribution of rms CDPP versus \emph{Kepler} magnitude is the minimum noise floor, with contributions from both photon noise, which is a pure Poisson noise and depends only on the magnitude, and the typical readout noise \citep{Christiansen2012PASP}. Thus, we determine the lower bound by applying a 4th order polynomial fit to the rrms CDPP at the bottom 0.5-th percentile of points within 0.2-magnitude-wide bins as a function of stellar apparent magnitude in the Gaia DR3 $G$ bandpass \citep[hereafter, $G$,][]{Jordi2010}. The polynomial function is given by
\begin{equation}\label{equ1}
\min(\log \mathrm{rrmsCDPP}_{k}) = p_0 G^4 + p_1 G^3 + p_2 G^2 + p_3 G + p_4
\end{equation}
where ${\rm rrmsCDPP_{k}}$ is the observed overall noise on the $k$-hr integration duration, $p_0$, $p_1$, $p_2$, $p_3$, and $p_4$ are the coefficients of the polynomial terms.

\figu{fig:cdpp06p5_vs_Gmag} illustrates an example of the polynomial fit to the entire sample of \emph{Kepler} targets with 6.5-hr integration duration. Given the primary range of stellar magnitude in \emph{Kepler} band is $Kp = 9\sim15$ \citep{Koch2010ApJ}, the fit applied for all \emph{Kepler} targets with $G$ in the range of $8\sim16$ which covers our sample stars of interest. We perform similar polynomial fits to determine the Poisson noise over 11 integration durations (i.e. [1.5, 2.5, 3.5, 5.0, 6.0, 6.5, 7.5, 9.0, 10.5, 12.0, 15.0] hr). The corresponding coefficients of the polynomial terms are listed in Table~\ref{tab:fit}. The polynomial relations are then subtracted in quadrature from rrms CDPP measures.

According to \citet{Gilliland2015AJ}, instrumental noise can be approximated by 13\% of squared rrms CDPP. We thus estimate the intrinsic stellar noise by the following form
\begin{equation}\label{equ2}
\begin{split}
    \mathrm{rrmsCDPP}_{\mathrm{intrinsic}}^{2} &= \mathrm{rrmsCDPP}^{2} \\
    &- (10^{\mathrm{min}(\log \, \mathrm{rrmsCDPP})})^{2} \\
    &-0.13 \, \mathrm{rrmsCDPP}^{2}.
\end{split}
\end{equation}
We note that taking a constant fractional value of 13\% as the instrumental noise is a simplified approximation, while a more realistic estimate for the \emph{Kepler} instrumental noise may need to consider possible variation with detector channels. For the current work, we do not expect such an approximation would cause a dramatic problem, given the relatively small contribution of this term compared to the photometric noise and the intrinsic noise. However, a further, detailed characterization of the $Kepler$ instrument noise will be very helpful for future studies.

\begin{figure}
\figurenum{2}
\epsscale{1.1}
\plotone{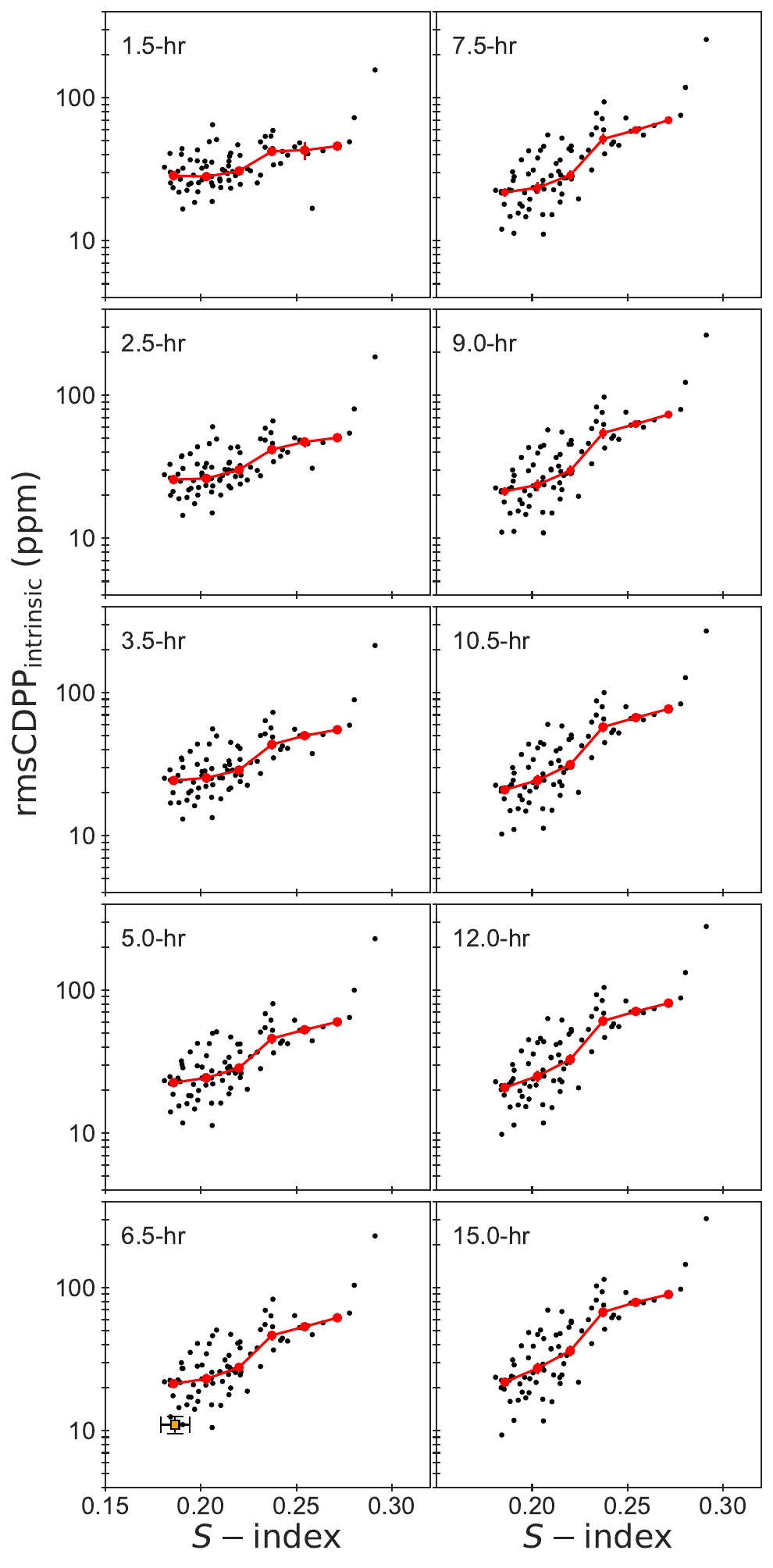}
\caption{\s{}-index versus intrinsic rrms CDPP on different integration duration for Sun-like stars with magnitudes in the range of $11.5<K_{p}<12.5$. The 6.5-hr rrms CDPPs are estimated from the accessible 6-hr rrms CDPPs that are divided by $(13/12)^{0.5}$ = 1.041. The red dot symbols connected with lines in each panel represent the median values of intrinsic rrms CDPP in 6 \s{}-index bins. The vertical line segments indicate standard errors of intrinsic rrms CDPP within the bin. The orange square with error bars in the bottom-left panel represents the Sun.}\label{fig:S_VS_rmscdpp}
\end{figure}

\begin{figure}
\figurenum{3}
\epsscale{1}
\plotone{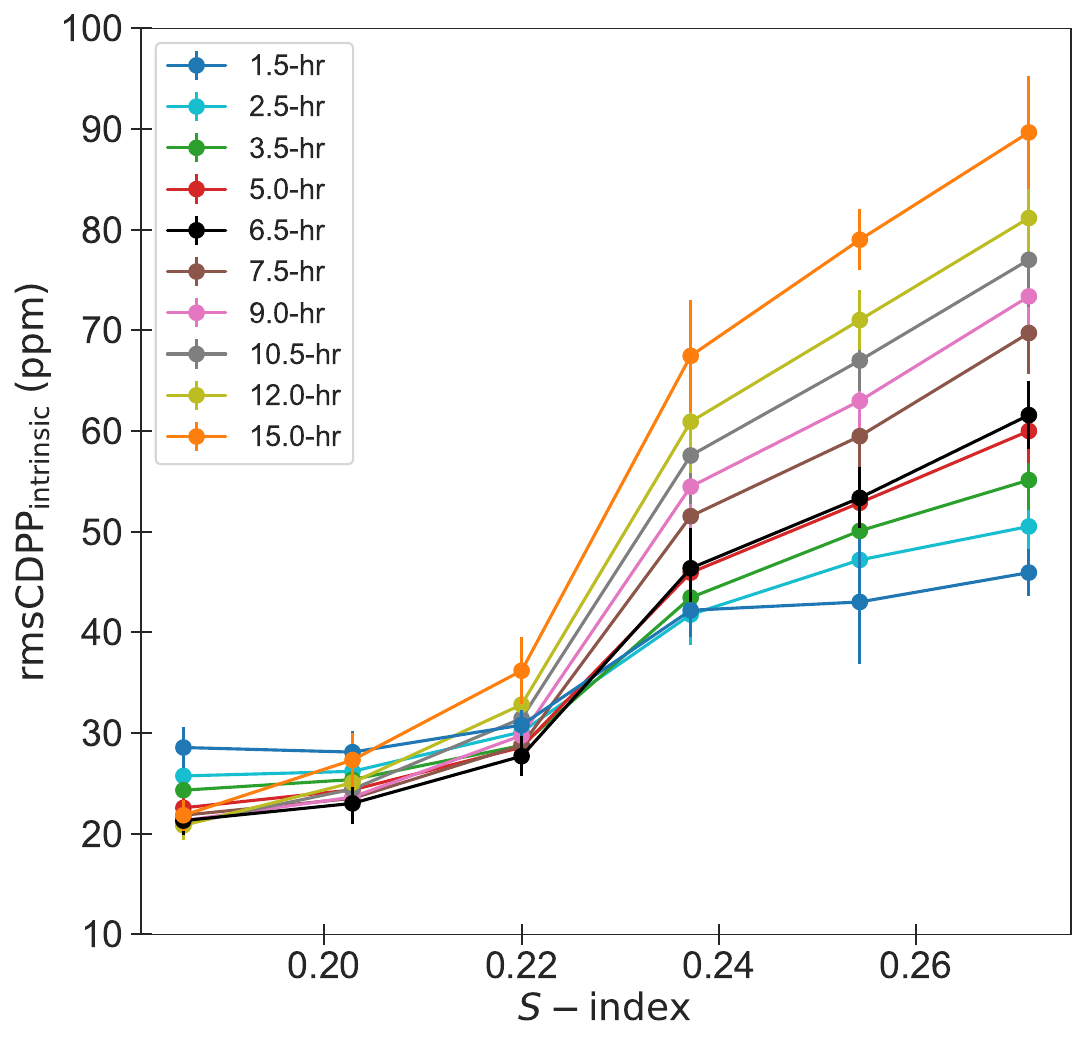}
\caption{The median values of intrinsic rrms CDPP as a function of average \s{}-index in 6 bins (see Section~\ref{sec:activity noise} for more details) for sun-like stars with magnitudes in the range of $11.5<K_{p}<12.5$. The error bars indicate standard errors of intrinsic rrms CDPP within the bins.
\label{fig:S_VS_rmscdpp_median}}
\end{figure}

\subsection{relation between intrinsic noise and stellar activity for solar-type stars} \label{sec:activity noise}
To explore the relation between stellar activity and intrinsic noise, we first focus on solar-type stars with \teff{}, \logg{} and \feh{} in the ranges 5677--5877 K, 4.34--4.54 and $-0.1$--0.1, respectively. We also limit the sample to a narrow magnitude range ($K_{p} = 11.5 \sim 12.5$) to avoid introducing dependence of noise on magnitude.

\figu{fig:S_VS_rmscdpp} shows the distribution of intrinsic rrms CDPP for different integration durations as a function of \s{}-index for the 77 selected solar-type stars. It clearly shows that the intrinsic rrms CDPP in each integration duration panel is positively correlated with \s{}-index. The Spearman’s rank order correlation coefficient $r_{s}$ are 0.45, 0.72, and 0.79 on 1.5-hr, 6.5-hr and 15.0-hr timescales, respectively. That is, the correlations are more significant for longer integration duration. The reason for this is the fact that behavior following from magnetic activity and rotation of solar-type stars are better elucidated at longer timescales \citep{Basri2013ApJ}. This result is compatible with \citet{Gilliland2011ApJS} who showed that the activity dominates at high stellar noise for solar-type stars based on the synthetic population. 

The bottom-left panel of \figu{fig:S_VS_rmscdpp} also shows the location of the Sun in the $S$-index and intrinsic noise plane. The solar\s{}-index is a mean of measurements by LAMOST during activity cycles 15–24 \citep{Zhang2020ApJ}, while the solar intrinsic noise is taken from \citet{Gilliland2011ApJS}. It can be seen that the Sun follows the same relation as other solar-type stars but has the lowest intrinsic noise and activity level, which might be a factor making it in place to hold a habitable planet.

To give a more explicit picture, we divide \s{}-index, whose values span from 0.16 to 0.28, into 6 bins and then derive the average \s{}-index as well as the median values of the intrinsic rrms CDPP in each bin for all the data sets shown in \figu{fig:S_VS_rmscdpp} (the red dots symbols). We further plot these median intrinsic rrms CDPP versus the average \s{}-index on different timescales in \figu{fig:S_VS_rmscdpp_median}. It can be found that the median intrinsic rrms CDPP on all timescales distribute mainly in the range of 20--30 ppm at the low-activity end, with the 1.5-hr duration slightly higher than others. Moreover, the dependency rises with increasing \s{}-index and becomes significant for stars with \s{}-index higher than $\sim0.22$. At the high-activity end, the noise levels on different timescales separate from each other from 40 to 90 ppm, with the 15-hr duration noise being twice that of 1.5-hr. This remarkable difference of intrinsic noise on the different duration timescales manifests the larger contribution induced by stellar activity to stellar noise on longer timescales, as we mentioned previously.

This result could provide valuable guidance to select targets for Earth-2.0 search in future missions. As a reference, the inactive stars with \s{}-index $< 0.22$, which corresponds to an intrinsic noise lower than $\sim30$ ppm, should be favored to increase the success rate of Earth 2.0 detection. 

We note the results are demonstrated based on solar-type stars with activity levels estimated from LAMOST low-resolution spectra. The \s{}-index value is subject to variation due to the resolving power of different instruments (as the lower the resolving power is the stronger the line cores are mixed with wings and, consequently, the larger are the Ca {\sc ii} H \& K fluxes). We inspect the influence of different resolving powers of instruments on our results in Sect.~\ref{sec:Discussion}. We collect stars in common with \s{}-index measured both from the LAMOST low-resolution spectra and other high-resolution spectra. After calibrating \s{}-index from the LAMOST scale to the conventional scale of Mount Wilson Observatory (hereafter, MWO) HK Project \citep{Wilson1978}, we find the increasing trends of stellar noise with activity levels are still significant except that the inflection point of $\textit{S}_{\rm MWO}$ extends slightly beyond 0.22 on the LAMOST scale. The median intrinsic rrms CDPP distribution below the inflection point is mainly concentrated in the 20--30 ppm range on timescales shorter than 12 hours (see Figure~\ref{fig:S_MWO_VS_rmscdpp_median}).

The dispersion shown in \figu{fig:S_VS_rmscdpp} could be caused by the intrinsic variability of stellar activities. For solar-type stars, the variations of \s{}-index values due to the intrinsic variability of activity are expected to be similar to those of the Sun (i.e., about 10\% from the mean value) \citep{Egeland2017}. On the other hand, the different ages of these stars could also lead to variations of the \s{}-index values. According to \citet{Zhang2020ApJ}, solar-type stars with near-solar rotation periods have chromospheric activities that are systematically higher than stars with undetected rotation periods, which, in one aspect, reflects the discrepancies in their ages. 

\begin{figure*}
\figurenum{4}
\includegraphics[scale=0.19]{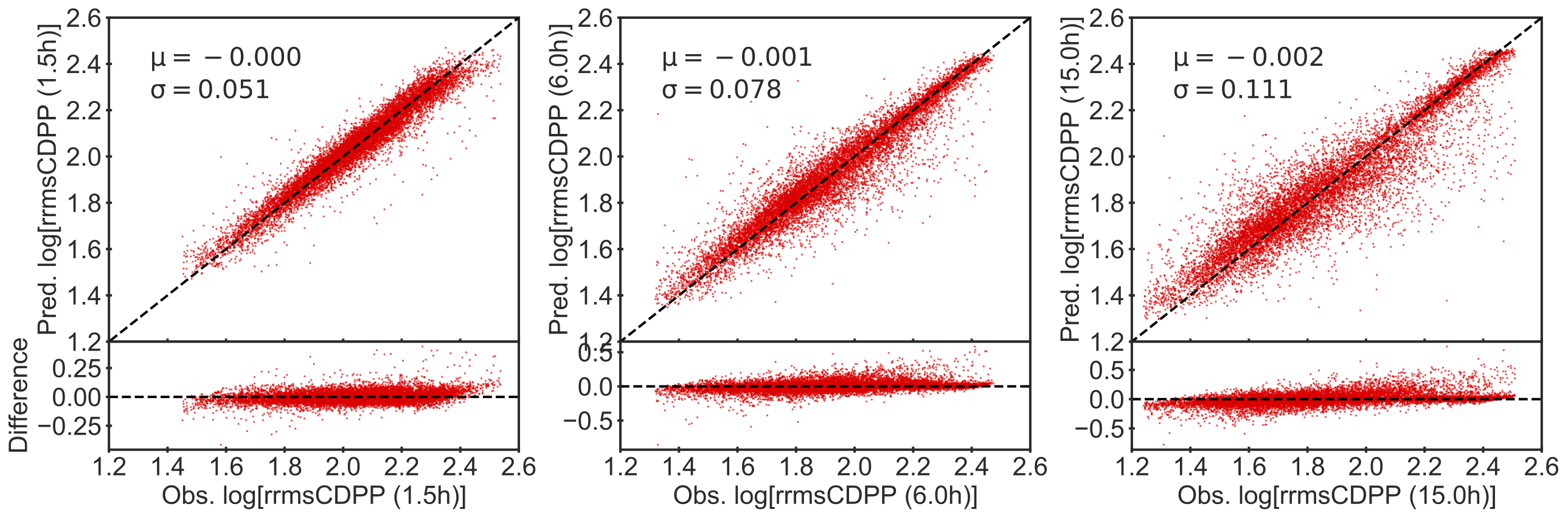}
\centering
\caption{Comparison of \textit{observed} with the \textit{predicted} rrms CDPP on 1.5-hr (left panels), 6.0-hr (middle panels), and 15.0-hr (right panels) transit scale, with the lower panels represent the respective residuals between \textit{observed} with the \textit{predicted} rrms CDPPs. The dashed lines represent perfect agreement.  The mean bias ($\mu$) with corresponding scatter ($\sigma$) is shown in the upper panels.
\label{fig:XGBoost_fit_rrmsCDPP}}
\end{figure*}

\section{predicting stellar noise for the LAMOST sample} \label{sec:fitting noise by ML}
\subsection{\texttt{XGBoost} regression algorithm} \label{sec:ML method}
Sect.~\ref{sec:activity noise} has shown the significant impact of stellar activity on intrinsic stellar noise by focusing on the solar-type star sample. The correlation between stellar activity and intrinsic noise is expected to be a ubiquitous effect for all stars with a radiative core and a convective envelope. Thus, we use the machine learning method to quantitatively characterize the relationship between the intrinsic stellar noise and stellar properties, i.e., stellar activity (\s{}-index), atmospheric parameters (\teff{}, \logg{}, \feh{}), and magnitude ($G$):
\begin{equation}
\resizebox{\linewidth}{!}{$
    \mathrm{rrmsCDPP}_{\mathrm{intrinsic}} =  f(\mathrm{s\text{-}index}, T_{\mathrm{eff}}, \log g, \mathrm{[Fe/H]}, \textit{G})
$}
\end{equation}
in which $T_{\rm eff}$ is used in its logarithmic form. This is important for target selections in transiting planet search missions if the noise of targets is known from these parameters. 

We predict individually the intrinsic rrms CDPPs on three representative timescales, i.e. 1.5-hr, 6.0-hr, and 15-hr. Since the 6.5-hr noise metrics are estimated by the accessible 6-hr noise metrics (see details in Section \ref{sec:data}), we predict the 6-hr noise metrics instead of 6.5-hr noise metrics. We also predict the observed overall rrms CDPPs using the same stellar properties for comparison. Note that the overall rrms CDPPs and intrinsic rrms CDPPs are predicted in their logarithmic form given they span several orders of magnitude.

\begin{figure*}
\figurenum{5}
\includegraphics[scale=0.19]{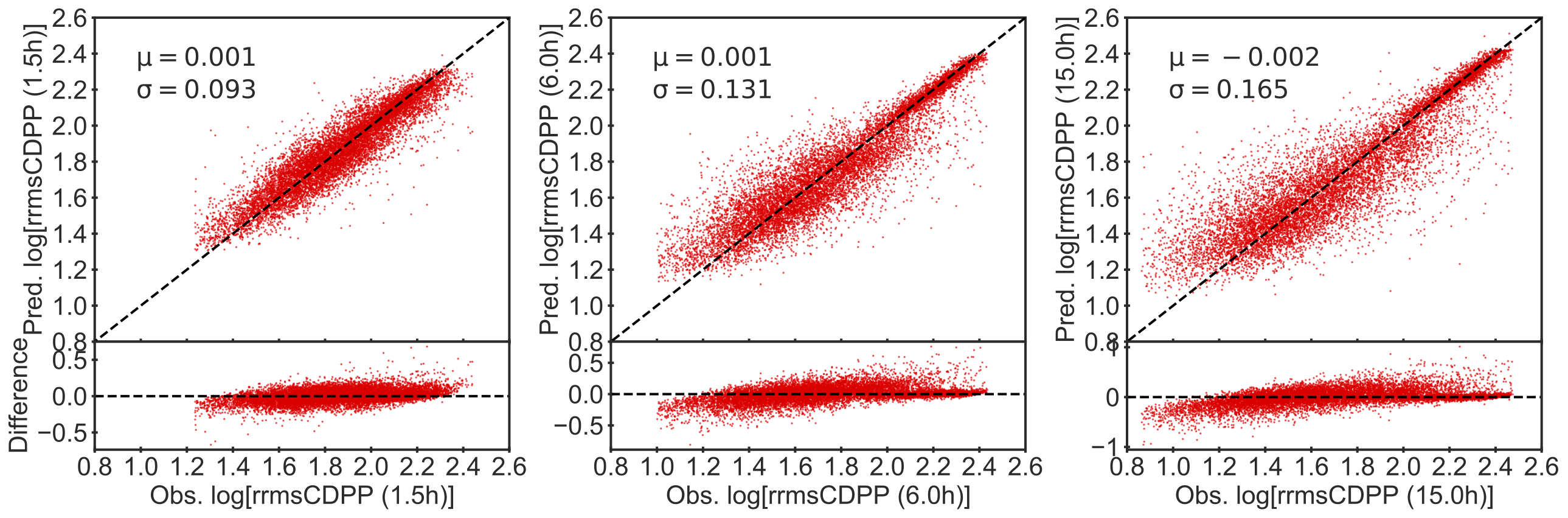}
\centering
\caption{The same as Figure~\ref{fig:XGBoost_fit_rrmsCDPP} but for \textit{intrinsic} rrms CDPP that removes Poisson noise as well as instrumental noise from \textit{observed} rrms CDPP using Equation~\ref{equ1} and Equation~\ref{equ2}.
\label{fig:XGBoost_fit_rrmsCDPP_i}}
\end{figure*}
For each noise metric, we apply the \texttt{XGBoost} regression algorithm to build a model. The \texttt{XGBoost} algorithm is an amelioration of the gradient boosting method, and it also leverages poor predictors by combining them together in a way that maximizes their predictive power \citep{Friedman2000, Friedman2001, FRIEDMAN2002, Chen2016}. It uses a more regularized model formalization to control overfitting and thus give better performance \citep{Chen2016}. 

We split the sample data into training and test sets with a ratio of 7:3. For this purpose, we eliminate stars with missing labels (i.e., input labels) and discard both the top and bottom 0.5-th percentile of data points for each label as the \texttt{XGBoost} algorithm is sensitive to outliers. For applying the \texttt{XGBoost} regression algorithm, we emprically set the optional hyperparameters, including the number of regression trees, the maximum depth, the learning rate, and the minimum child weight. To obtain the best set of values for these hyperparameters, we perform a 5-fold cross-validated grid-search using the \texttt{scikit-learn} GridSearchCV routine \citep{Pedregosa2011}. We tune these to best fit our data without over- or underfitting by using the calculated coefficient of determination of the predictions, i.e. $R^{2}$ score, as evaluation metrics. The $R^{2}$ score is the proportion of the variance in the dependent variable that is predicted from the independent variable. It generally ranges from 0 to 1, which indicates the level of variation in the given data set or indicates the accuracy of the prediction on average. The estimated $R^{2}$ over $n_{\rm samples}$ is defined as
\begin{equation} 
R^2 = 1 - \frac{\sum_{i=1}^{n_{\rm samples}} (y_i - \hat{y}_i)^2}{\sum_{i=1}^{n_{\rm samples}} (y_i - \bar{y})^2} 
\end{equation}
where $y_i$ is the observed value of the $i$-th sample, $\hat{y}_i$ is the corresponding predicted value, and $\bar{y}$ is the mean of the observed values.
After training the algorithms to predict observed overall noise and intrinsic stellar noise in the training sets, we apply these trained algorithms to the testing set and obtain a list of corresponding predicted noise metrics. 

\subsection{examination of model performance with test set}\label{sec:ML results}
Figure~\ref{fig:XGBoost_fit_rrmsCDPP} illustrates the comparison between the observed and predicted rrms CDPP for the test set on 1.5-hr, 6.0-hr, and 15.0-hr transit scale, individually. The mean bias ($\mu$) with corresponding scatter ($\sigma$) are calculated and given in corresponding panels. The scatter of prediction for overall noise is 0.078 on 6.0-hr timescale, equivalently to 17 ppm, indicating that \s{}-index, \teff{}, \logg{}, \feh{}, and $G$ could be successfully used to predict the observed noise to such a level. Stars with noise on 1.5-hr, 6.0-hr, and 15.0-hr timescales have $R^{2}$ scores of 0.93, 0.90, and 0.85, respectively, which means a smaller difference and higher precision of the prediction for shorter integration duration.

Figure~\ref{fig:XGBoost_fit_rrmsCDPP_i} illustrates the comparison between the observed and predicted {\it intrinsic} rrmsCDPP on 1.5-hr, 6.0-hr, and 15.0-hr transit scale, individually. The comparison for the intrinsic noise returns a scatter of 0.131 on 6.0-hr timescale, equivalently to 19 ppm, with the separated scatter for the giants (\logg{}$\le$4) and dwarfs (\logg{}$>$4) being 16 and 21 ppm, respectively. Stars with intrinsic noise on 1.5-hr, 6.0-hr, and 15.0-hr timescales have $R^{2}$ scores of 0.85, 0.84, and 0.81, respectively. Compared to the case of overall rrmsCDPP prediction (Figure~\ref{fig:XGBoost_fit_rrmsCDPP}), the scatter in the intrinsic noise prediction is slightly increased (19~ppm versus 17~ppm), and the $R^{2}$ scores are slightly decreased. We believe this is mainly due to uncertainties induced in the process of extracting intrinsic stellar noise from the overall rrmsCDPP. 

\begin{figure}
\figurenum{6}
\includegraphics[scale=0.35]{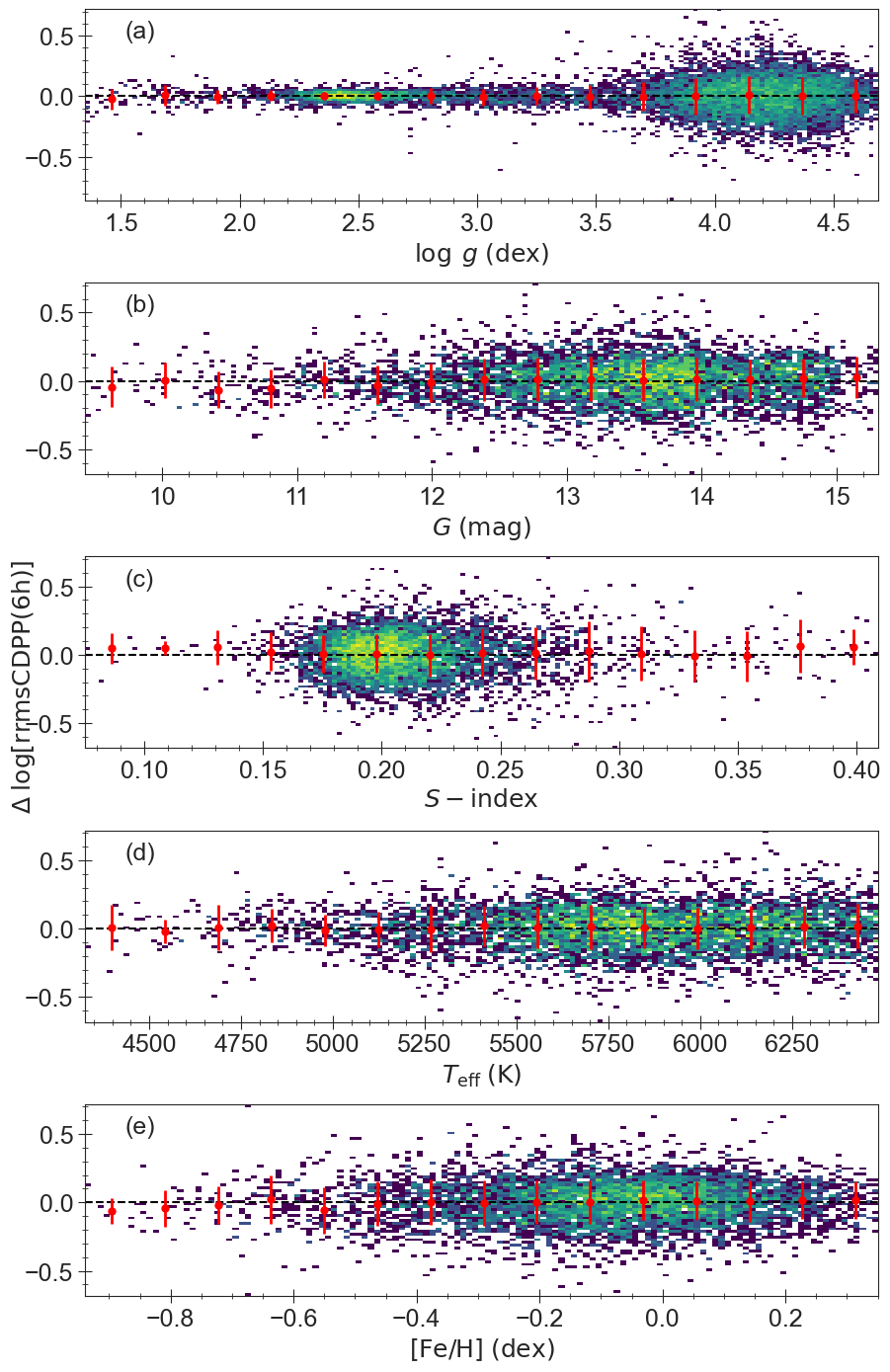}
\centering
\caption{Residuals between \textit{intrinsic} and \textit{predicted} rrmsCDPP on 6.0-hr as functions of individual labels for the whole (dwarfs and giants) test set (panel a) and for the dwarfs (panel b, c, d, and e), color-coded by the stellar number density. The dashed lines represent perfect agreement. Red dots and vertical error bars are the medians and standard deviations of the residuals within 15 bins for each label.
\label{fig:residual_params}}
\end{figure}

Figure~\ref{fig:residual_params} shows the residuals between intrinsic and predicted rrms CDPP on 6.0-hr timescale as a function of the values of the label. According to label importance built in the \texttt{XGBoost} algorithm, we arrange the panels in descending order based on the contribution of each label to the intrinsic noise. The intrinsic noise is mostly related to \logg, as the mean noise level decreases from 107~ppm for giants to 46~ppm for dwarfs in our sample. However, as shown in panel (a), the residuals of the \texttt{XGBoost} prediction are significantly small for giants. As our major object is dwarf star, we only display the residuals for dwarfs with \logg{} $>$ 4\,dex in the other panels ($b$--$e$). The Figure illustrates that our model has achieved a good prediction for the intrinsic noise of dwarfs without significant bias to stellar labels.

In order to quantify how our results are affected by measurement errors in the input labels, we perform Monte-Carlo experiments for the test set. For each star in the test set, we randomly draw 500 sets of labels from Gaussian distributions, centered on their measured values with a dispersion equalling their measurement errors. We re-predict the intrinsic rrmsCDPP for the 6.0-hr case by using the optimal \texttt{XGBoost} model trained above. The scatter of the 500 rrmsCDPP predictions is then calculated for each star. The median value of the scatters for the whole test set is 15~ppm (0.106~dex), and is 12~ppm for giants (\logg{}$\le$4), 17~ppm for dwarfs (\logg{}$>$4). 

Given the total intrinsic noise is $\sim19$~ppm for dwarfs, while the intrinsic noise arised from uncertainties in the stellar labels is $\sim15$~ppm, the root of their squared difference suggest an extra component of $\sim12$~ppm, which is likely contributed by uncertainties in the intrinsic rrmsCDPP estimates. This also means that the underlying relation between the rrmsCDPP noise and the stellar labels are rather tight, as any intrinsic scatter, if exists, should be smaller than 12~ppm.

To quantify possible statistic fluctuation uncertainty in the \texttt{XGBoost} modeling process, we also implement a bootstrap experiment. We re-sample the training set randomly from the original training sample 50 times. We then obtain 50 \texttt{XGBoost} models from these training sets, and derive their rrmsCDPP predictions for the test set. The scatters among the 50 rrmsCDPP predictions for individual in the test set has a median value of 0.024~dex, indicating that statistic uncertainty arising from the \texttt{XGBoost} modeling is negligible. This is not surprising as the size of the training sample is sufficiently largefor our modelling purpose.

\begin{deluxetable*}{cccccccccccc}
\tiny
\caption{Sample entries of deduced photometric noise of the 1,358,275 stars. \label{tab:sample}}
\tablehead{
$Gaia\,\rm DR3$ & $G$ & $T_{\rm eff}$ & $\log$ g & [Fe/H] & $\emph{S}$ & $\rm CDPP_{1.5h}$ & $\rm CDPP_{6.0h}$ & $\rm CDPP_{15.0h}$ & $\rm CDPP_{i,1.5h}$ & $\rm CDPP_{i,6.0h}$ &
$\rm CDPP_{i,15.0h}$ }
\startdata
3385982517716355584& 14.00& 4833.43& 3.30& 0.07& 0.1289& 2.2599& 2.0263& 1.8805&  2.1325& 1.9167& 1.7862\\
3384387469942051712& 13.50& 4890.12& 2.37& -0.23& 0.1251& 2.2647& 2.3599& 2.4082& 2.1827& 2.3339& 2.3627\\
3385677094001877504& 14.29& 6159.51& 4.05& -0.22& 0.1847& 2.1696& 1.9285& 1.8049& 1.9405& 1.6761& 1.5771\\
3385868164209420160& 14.36& 5587.70& 4.32& -0.24& 0.2073& 2.2055& 1.9450& 1.7978& 1.9763& 1.7231& 1.5566\\
3384171347187569536& 13.85& 4911.19& 2.42& -0.29& 0.1530& 2.3254& 2.4239& 2.4101& 2.2453& 2.3450& 2.3874\\
...&...&...&...&...&...&...&...&...&...&...&...\\
\enddata
\tablecomments{}{The generated overall noise listed in columns 7-9 as well as intrinsic noise listed in columns 10-12 are given in their logarithmic form.}
{(This table is available in its entirety in the machine-readable form.)}
\end{deluxetable*}
\subsection{generating stellar noise for the LAMOST sample}
We have so far achieved a series of training models that enable predictions for photometric noise on different timescales of stars that have parameters falling within the trained parameter space. For the generalization set, we start with the catalog of the A, F, G, and K type star from the LAMOST DR7 (v1.1) low-resolution survey which includes 6,199,917 spectra \footnote{http://dr7.lamost.org/v1.1/catalogue}. The \s{}-indexes are then determined for 2,278,792 spectra with S/Ns greater than 50 in the catalog. To obtain the unique source, we use the CDS X-match service in TOPCAT \citep{Taylor2005} to consider the epoch of Gaia DR3 stars. Targets are identified with R.A. and Dec. coordinates within 2.0 arcsec, 1,763,868 stars are cross-matched. To maintain consistency with the source of the atmospheric parameters, we cross-match these stars with the catalog of stellar parameters of 7 million stars from the LAMOST DR7 spectra based on the data-driven Payne (see Section~\ref{sec:data} for more details). It is of crucial importance to ensure that the generalization set parameter space overlaps with our training set as much as possible. Hence, we further select a generalization set with the values of their labels (i.e. \teff{}, \logg{}, \feh{}, $G$, and \s{}-index) fall within the maximum and minimum values of the corresponding labels in the training set, 1,358,275 stars eventually remain. Table \ref{tab:sample} lists the generalization set along with labels as well as their predicted photometric noises (including overall noise and intrinsic noise) on different timescales for 1,358,275 targets in the LAMOST field.

\begin{figure}
\figurenum{7}
\epsscale{1.1}
\plotone{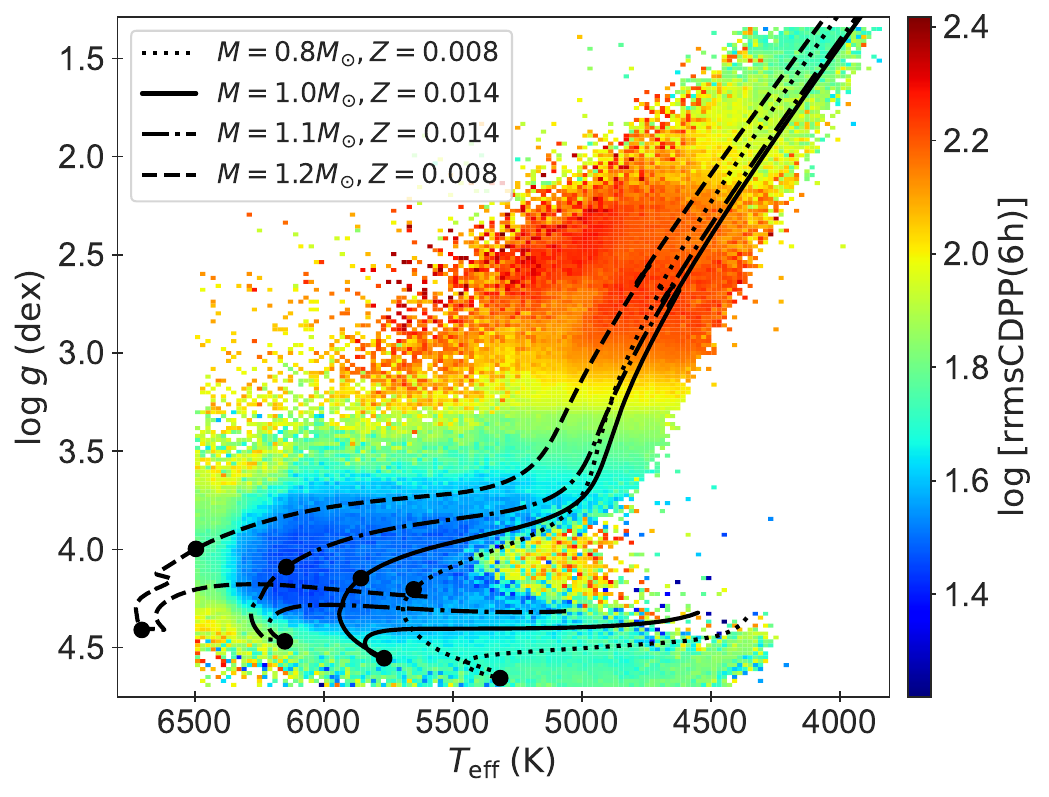}
\caption{Distribution of 6-hr intrinsic rrms CDPP values for the generalization set in the \teff{}-\logg{} plane, binned by 20 K by 0.02 dex and color-coded by the median value of rrms CDPP on a logarithmic scale. The black lines show evolutionary tracks from PARSEC v2.0 \citep{Nguyen2022}, with the mass and metallicity shown in the panel. The two black dots along each track indicate the start point of main-sequence as well as sub-giant phases. \label{fig:gener_rrmsCDPP06p0_i_HRD}}
\end{figure}

\figu{fig:gener_rrmsCDPP06p0_i_HRD} shows distributions of generated 6-hr intrinsic rrms CDPP in \teff{}-\logg{} space, together with the model tracks from PARSEC \citep{Nguyen2022}. It is clear that the noise levels vary with stellar evolutionary phases. Regarding the stars with relatively low noise levels, like $\rm log\;rrms CDPP < 1.6$, which will be the special interest targets in the transiting-planet surveys, we found that part of them have ended their main-sequence phase and entered sub-giant phase. The magnetic field wanes on sun-like stars as they evolve off the main-sequence, which leads to weak activity levels and negligible contributions to intrinsic stellar noise. Although these evolved stars are promising targets according to
their low activity level, their internal structures and surface properties are unstable because of rapid evolutionary changes, leading to inhospitable to life. Moreover, the orbital distance that corresponds to the habitable zone, which is based on stellar irradiance and the host star’s SED, moves outward with increasing stellar luminosity during a star’s evolution \citep[e.g.][]{Rushby2013, Luger2015, Ramirez2014, Ramirez2016}, which decreases the probability that transiting by terrestrial planets in front of these evolved stars in light curves \citep{Ramirez2016}. 

For the stars on main-sequence, we found the noise levels tend to drop off over time, which is, in a way, a good manifestation of the evolution of stellar activity and has been demonstrated in previous studies \citep[e.g.,][]{Chen2021AJ, Ye2024ApJS}. Stellar magnetic activity is related to stellar rotation but also causes the star to lose angular momentum over time via braking from a magnetic wind, which offers the promise that one might be able to trace stellar ages with activity levels \citep[See][for a review]{Brun2017}. In general, those prioritizing targets with low intrinsic stellar noise for transiting planet surveys would have evolved for a long lifetime. In addition, we see that cool stars have higher noise levels than hot stars, which is commonly interpreted as an indication of the high activity levels of cool stars. On the right area of the panel, there is a small number of relatively high-level-noise stars which may be misplaced due to the large error of \logg{} and they are supposed to be located at the lower-luminosity region. 

The bulk of less-luminous red giants are generally the noisiest. In this situation, there should be little probability that transiting by terrestrial planets in front of these evolved stars can be detected with achievable precision of current known missions. In addition, the associated depth in the light curve varies as the squared ratio between planet and star radii, $(R_{p}/R_{\star})^2$ \citep{Heller2019}, the giant stars with expanding out layers would lead to indiscernible depth in light curves arising from potential Earth-sized planets. On the contrary, the noise of stars with \logg\,$<$\,2\,dex are relatively lower. Since the noise metrics in the \emph{Kepler} light curves are determined by decomposing the data in the time-frequency domain \citep{Jenkins2002ApJ}, the low-frequency pulsation in a longer period regime produced by these luminous red giants are usually beyond the High-Pass Filter within several-hour time scale, which leads to the underestimated noise level for these targets. 

\section{Discussion}\label{sec:Discussion}

Our results have shown that there exists a tight correlation between the stellar intrinsic photometric noise and the chromospheric activity ($S$-index), and this relation can be used to predict the intrinsic photometric noise from spectroscopic $S$-index and stellar atmospheric parameters, which is valuable for exoplanet detection. An important factor that affects the robustness of our method is the $S$-index measurement. While our analysis above has shown that the internal error of the LAMOST $S$-index measurement for our sample stars is small ($<0.02$), an external comparison with literature high-resolution spectra is insightful for revealing any systematic uncertainties in our results.

\subsection{uncertainties due to stellar activity measurements}
\begin{figure*}
\figurenum{8}
\epsscale{1.1}
\plotone{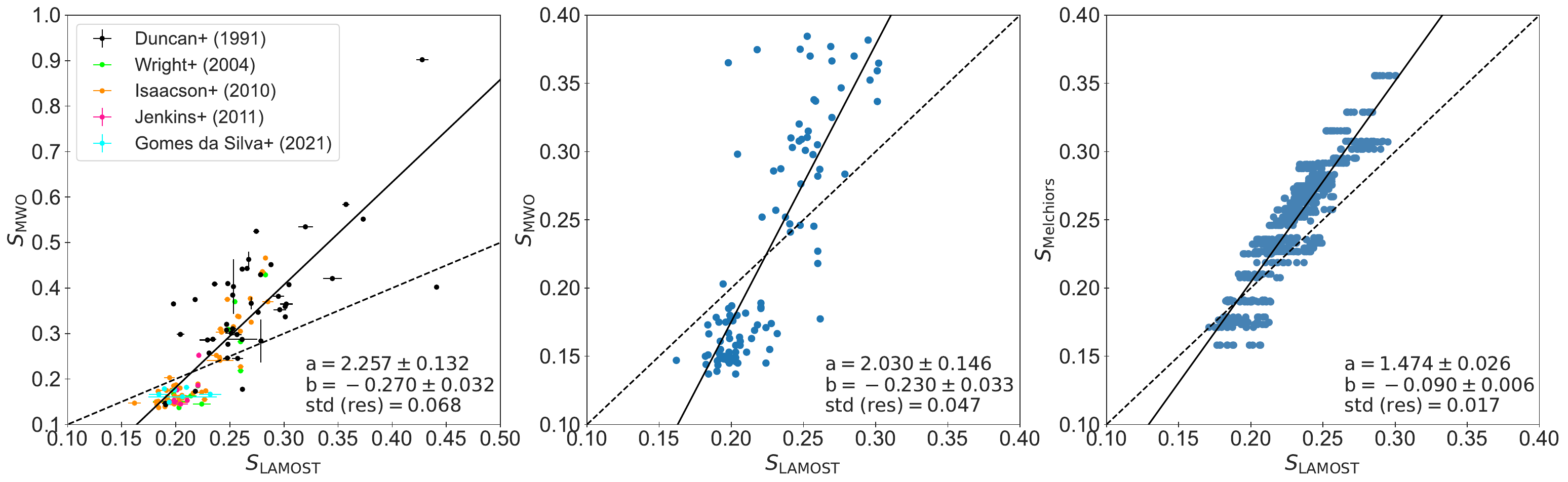}
\caption{Left: Calibration of \s{}-index from the LAMOST to the MWO using 124 stars within $5400<T_{\rm eff}<6500$~K and $\textit{S}_{\rm MWO}$ values from literatures shown in the plot. Error bars are only shown for the LAMOST measurements, as errors for literature $\textit{S}_{\rm MWO}$ are not available. In many cases, the error bars are smaller than the symbol size. Middle: Same as the left panel but for stars with $S_{\rm MWO}<0.4$. Right: Comparison of \s{}-index measured from high-resolution spectra for 56 dwarf stars with that measured from the same spectra but degraded to low resolution power. Each of the high-resolution spectra are degraded to 11 low-resolution spectra, for which the resolution power is adopted to be the LAMOST resolution but added by a random offset within $\pm20$~per cent (see text). The dashed lines in all the panels delineate the 1:1 line. The solid black lines represent the best linear fit with slope (a), intercept (b), and standard deviations of the fitting residuals marked in the plot.}\label{fig:S_scale}
\end{figure*}

We collect 186 stars that have \s{}-index from both the LAMSOT low-resolution spectra and literature high-resolution spectra (\citet[][73 stars]{Duncan1991}, \citet[][17 stars]{Wright2004}, \citet[][74 stars]{Isaacson2010}, \citet[][11 stars]{Jenkins2011aa}, \citet[][11 stars]{Silva2021}). All the \s{}-index values in the latter have been calibrated to the MWO measurement. The effective temperatures of the collected stars span from 3929~K to 7392~K. For our purpose, we limit the analysis to stars with $5400<T_{\rm eff}<6500$~K and $\log~g>3$, which leaves 124 stars. The left panel of Figure~\ref{fig:S_scale} shows a comparison of \s{}-index between LAMOST and literature values, while the middle panel shows the comparison particularly focused on stars with lower activity.

The LAMOST measurements exhibit a good consistency with literature values, but there exists a systematic trend that deviates from the 1:1 line. A linear fit to the trend yields 
\begin{equation}
S_{\rm MWO} = 2.257(\pm0.132)\cdot S_{\rm LAMOST}-0.270(\pm0.032).
\end{equation} 
Such a systematic trend is a consequence of the different spectral resolution for the $S$-index measurements. Beyond the trend, there also exists a star-to-star scatter concerning the linear fit, which is 0.068 for the overall sample, and 0.047 for stars with $S_{\rm MWO}<0.4$. We believe such a scatter is mainly caused by intrinsic temporal variations of the stellar activity levels, as the LAMOST and literature spectra were taken at different epochs spreading decades.

To validate this speculation, we made an independent test by directly degrading the high-resolution spectra to the LAMOST resolution. We adopt the MELCHIORS database \citep[$R=85,000$;][]{Royer2024}, from which we selected 56 spectra of F/G/K type dwarf stars with $5400<T_{\rm eff}<6500$~K, based on stellar parameters derived from Gaia BP/RP spectra \citep{2023A&A...674A...2D, 2021A&A...652A..86C} with the {\sc DD-Payne} (Xiang et al. in prep.). For each high-resolution spectrum, we degrade it to 11 low-resolution spectra that have different resolution powers with random values within $\pm$20\% of the mean resolution of LAMOST, mimicking the fiber-to-fiber variation of the LAMOST spectral resolution \citep[e.g.][]{Xiang2015}.

The right panel of Figure~\ref{fig:S_scale} presents a comparison of \s{}-index measured from the MELCHIORS high-resolution spectra and that from spectra degraded to LAMOST resolution. It shows a similar systematic trend to the left and middle panels, validating that the systematic trend is mainly a consequence of different resolutions. However, the star-to-star scatter is only 0.017, which is consistent with measurement errors of the $S$-index (Sect.~\ref{sec:data}) but much smaller than the middle panel, verifying the star-to-star scatter of $S$-index between LAMOST and literature shown in the left and middle panels are likely due to intrinsic temporal variations of stellar activity.

For solar-type stars, an extra uncertainty in $S$-index of $\sim0.047$ due to temporal variations of stellar activity will cause a median uncertainty of $15\,$ppm in the rrmsCDPP for the 6.0-hr case. Considering the $Kepler$ photometric observations and the LAMOST spectroscopic observations used in this study were implemented at similar epochs, with a difference of $\lesssim5$ years in typical, we expect the effect due to temporal variation of stellar activity is insignificant for our rrmsCDPP prediction, except for young and active stars with rapid activity variation. This, in
turn, also implies that in order to have a good estimate of the stellar intrinsic noise with our method for future planet detection surveys, it is necessary to make sure the stellar activity is measured from spectra taken at a similar epoch.

\subsection{stellar activity -- intrinsic photometric noise relation in MWO scale}
\begin{figure}
\figurenum{9}
\epsscale{1}
\plotone{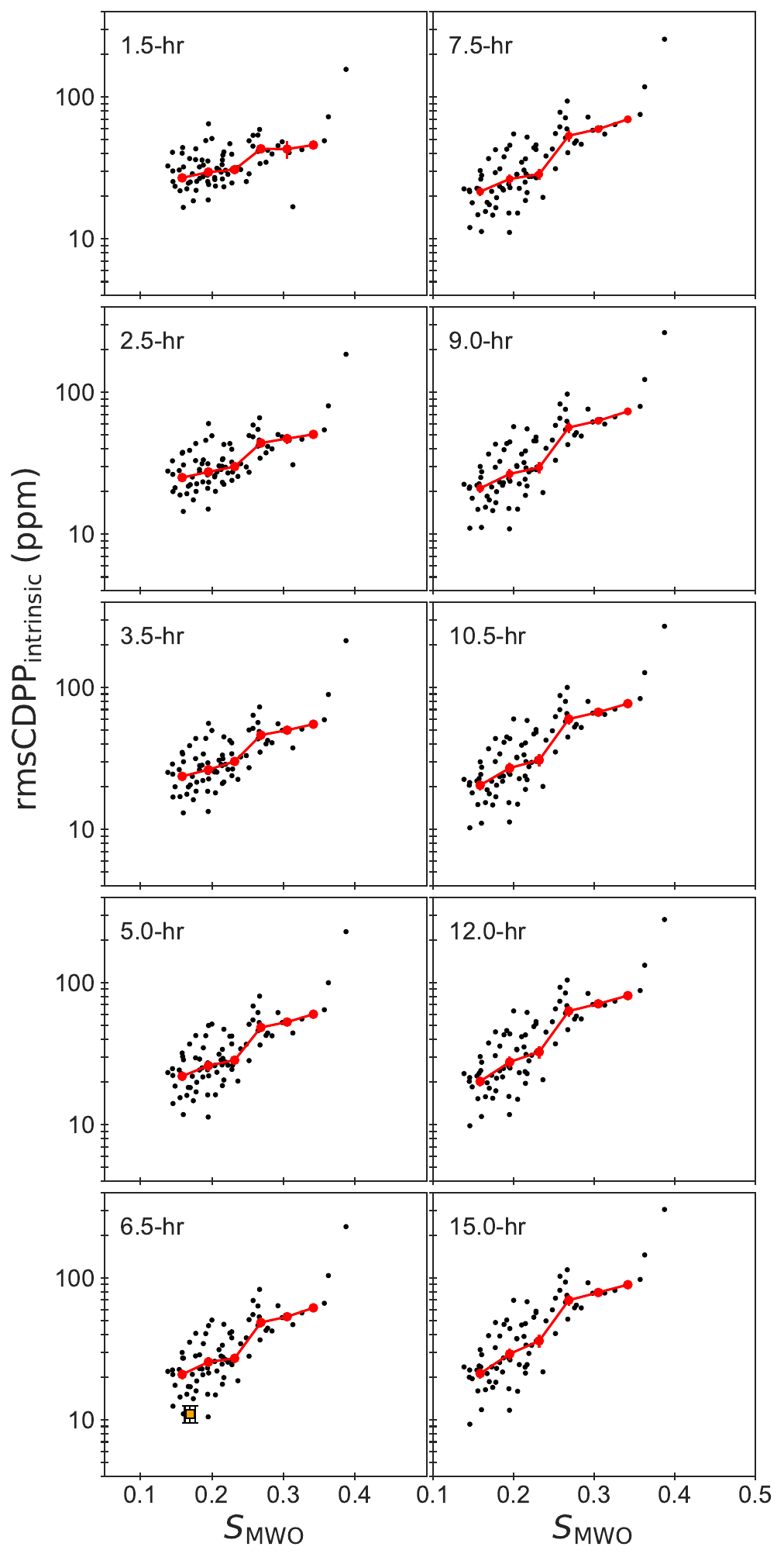}
\caption{The same as Figure.~\ref{fig:S_VS_rmscdpp} but using calibrated \s{}-index on the MWO. The orange square with error bars represents the Sun with S-index
value on the MWO during activity cycles 15–24 from \citep{Egeland2017} and intrinsic solar noise value from \citep{Gilliland2011ApJS}. \label{fig:S_MWO_VS_rmscdpp}}
\end{figure}

\begin{figure}
\figurenum{10}
\epsscale{1.0}
\plotone{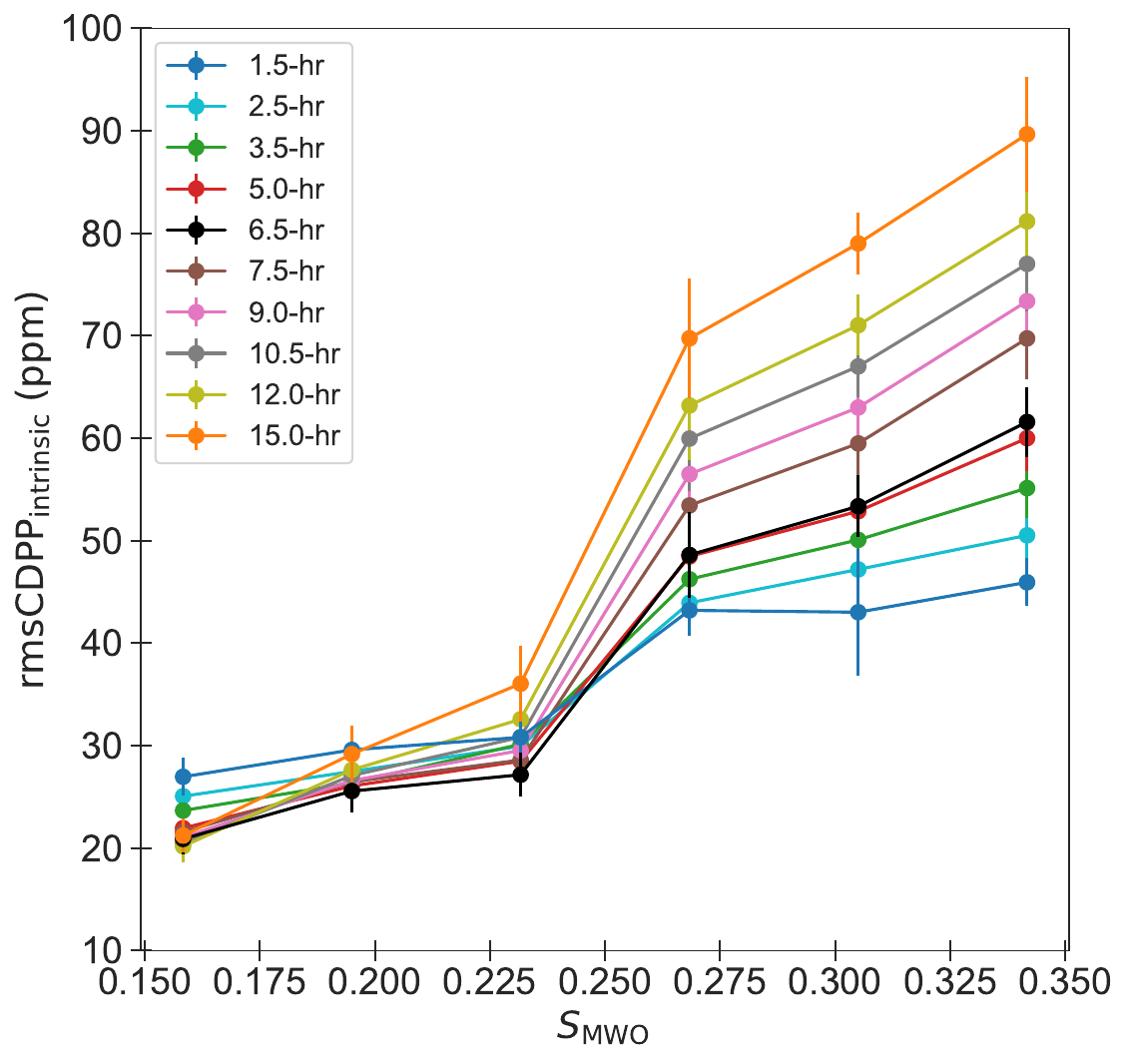}
\caption{The same as Figs.~\ref{fig:S_VS_rmscdpp_median} from the main text but using calibrated \s{}-index on the MWO.
\label{fig:S_MWO_VS_rmscdpp_median}}
\end{figure}

Irrespective of the temporal variation of stellar activity, we have repeated Figs.~\ref{fig:S_VS_rmscdpp}~ and ~\ref{fig:S_VS_rmscdpp_median} but for $S$-index calibrated to the MWO scale.  Figure~\ref{fig:S_MWO_VS_rmscdpp} shows that the positive relations between \s{}-index on the MWO scale and intrinsic rrmsCDPP are tenable on each integration duration panel. As illustrated by Figure~\ref{fig:S_MWO_VS_rmscdpp_median}, typical values of the intrinsic rrmsCDPP are 20--30~ppm for solar-type stars with \s{}-index lower than $~0.22$. Beyond this inflection point, the positive relations become significant. These results are accordant with that using \s{}-index derived from the LAMOST spectra (Figure~3).

\section{Conclusion} \label{sec:conclusions}
Stellar intrinsic photometric noise arising from magnetic activity is a main interference in detecting exoplanets using transit methods. In this work, we have investigated the relation between the stellar intrinsic photometric noise, as quantified by the $Kepler$ rrmsCDPP, and the stellar chromospheric activity \s{}-index derived from the LAMOST survey spectra. Our results revealed that, for solar-type stars, there exists a clear positive correlation between $S$-index and rrmsCDPP. Inactive stars with \s{}-index lower than $\sim0.22$ mainly possess low intrinsic noise, with rrmsCDPP values of 20--30 ppm, while the intrinsic noise increases dramatically for more active stars with $S$-index higher than 0.22. The correlation also shows a clear dependence on the photometric integration duration, as it becomes stronger for longer integration duration.

We then have built an empirical relation between the intrinsic noise and the stellar labels, including the \s{}-index, \teff{}, \logg{}, \feh{}, and apparent magnitude, using the  \texttt{XGBoost} regression algorithm. Internal and external examinations suggest the relation is robust, and our approach has achieved a typical precision of 20~ppm for inferring the intrinsic noise from the \s{}-index and other stellar labels on a 6-hour integration duration. We have applied this empirical relation to the full LAMOST spectra database, and obtained the intrinsic noise predictions for 1,358,275 stars. The resultant catalog is publicly available and expected to be valuable for optimizing target selection for future exoplanet-hunting space missions, such as the Earth 2.0 mission.

\vspace{7mm} \noindent {\bf Acknowledgments}
This work is supported by the National Natural Science Foundation of China under grant No.12103063. M.X. acknowledges financial support from NSFC Grant No.2022000083. and National Key R\&D Program of China Grant No. 2022YFF0504200. Jian Ge, Hui Zhang and Jiwei Xie acknowledges financial support from the Strategic Priority Program on Space Science of Chinese Academy of Sciences under grant No. XDA15020600. H.Z. acknowledges financial support from NSFC Grant No.12073010. This work is also supported by the Joint Research Fund in Astronomy (U2031203) under cooperative agreement between the National Natural Science Foundation of China (NSFC) and Chinese Academy of Sciences (CAS), and NSFC grants(12090040, 12090042). Y.W. acknowledges financial support from NSFC Grant No.12103064. Hong-Liang Yan acknowledges financial support from NSFC Grant No.12022304, 12373036, 12090044, and support from the Youth Innovation Promotion Association of the Chinese Academy of Sciences. This work is partially supported by the CSST project.
We acknowledge the entire \emph{Kepler} team and everyone involved in the \emph{Kepler} mission. Funding for the \emph{Kepler} Mission is provided by NASA’s Science Mission Directorate.
Guoshoujing Telescope (the Large Sky Area Multi-Object Fiber Spectroscopic Telescope, LAMOST) is a National Major Scientific Project built by the Chinese Academy of Sciences.
Funding for the project has been provided by the National Development and Reform Commission. LAMOST is operated and managed by the National Astronomical Observatories, Chinese Academy of Sciences.This work has made use of data from the European Space Agency (ESA) mission Gaia (\url{https://www.cosmos.esa.int/gaia}), processed by the Gaia Data Processing and Analysis Consortium (DPAC, \url{https://www.cosmos.esa.int/web/gaia/dpac/consortium}). Funding for the DPAC has been provided by national institutions, in particular the institutions participating in the Gaia Multilateral Agreement.



\clearpage

\bibliography{ref.bib}

\label{lastpage}

\end{document}